\documentclass[11pt]{article}
\usepackage[english]{babel}
\usepackage[latin1]{inputenc}
\usepackage{latexsym,amsfonts,amsmath,amssymb,euscript,epsfig,dsfont,graphicx,color,float}
\usepackage[active]{srcltx}
\usepackage[T1]{fontenc}
\DeclareMathOperator{\co}{co}
\DeclareMathOperator{\NC}{NC}
\DeclareMathOperator{\LIP}{Lip}
\DeclareMathOperator{\Chi}{\mathcal{X}}

\DeclareMathOperator{\Id}{Id}
\newcommand{\cS}{\mathcal{S}}
\newcommand{\R}{\mathbb{R}}

\title{On an unified framework for approachability in games with or without signals
}

\author{Vianney Perchet \thanks{Laboratoire de Probabilit\'es et de Mod\`eles Al\'eatoires, Universit\'e Paris 7, 175 rue du Chevaleret, 75013 Paris. vianney.perchet@normalesup.org} and M. Quincampoix\thanks{Laboratoire de Math\'ematiques, UMR6205, Universit\'e de
Bretagne Occidentale, 6 Avenue Le Gorgeu, 29200 Brest, France}}

\begin{document}
\maketitle
\newcounter{compteur}
 \newcounter{hypothese}
 \newcounter{rem}
 \newcounter{def}
 \newcounter{exa}
\newtheorem{proposition}[compteur]{Proposition}
\newtheorem{theorem}[compteur]{Theorem}
\newtheorem{lemma}[compteur]{Lemma}
\newtheorem{corollary}[compteur]{Corollary}
\newtheorem{hypo}[hypothese]{Assumption}
\newtheorem{definition}[def]{Definition}
\newtheorem{remark}[rem]{Remark}
\newtheorem{example}[exa]{Example}

\begin{abstract}
We unify standard frameworks for approachability both in full or partial monitoring by defining a new abstract game, called {\em the purely informative game}, where the outcome at each stage is the maximal information players can obtain, represented as some probability measure. Objectives of players can be rewritten as the convergence (to some given set) of sequences of averages of these probability measures. We obtain new results extending the approachability theory developed by Blackwell moreover
this new abstract framework enables us to  characterize approachable sets with, as usual, a remarkably simple and clear reformulation for convex sets.

Translated into the original games, those results become the first necessary and sufficient condition under which an arbitrary set is approachable and they cover and extend previous known results for convex sets.
We also investigate a specific class of games where, thanks to  some unusual definition of averages and convexity, we again obtain a complete characterization of approachable sets along with rates of convergence.
\end{abstract}

{\bf \large Introduction} $\;$
Repeated games can be studied by considering sequences of payoffs
and  constructing, stage by stage,  strategies with the requirement
that the outcome at the next stage will have  good properties given
the past ones. Perhaps, the most revealing examples of this claim
are  Shapley's~\cite{Shapley} operator that describes the
value of  stochastic zero-sum games,  the  {\em exponential weight algorithm} for predictions
with expert advices (see e.g.\ Cesa-Bianchi and
Lugosi~\cite{CesaBianchiLugosi}, Chapter 6) or
Blackwell's~\cite{BlackwellAnalogue} approachability theory.

We recall that in a two-person  repeated game with vector payoffs in
some euclidian space $\mathds{R}^k$, a player can {\em approach} a given
set $E \subset \mathds{R}^k$, if he can insure that, after some
stage and with a great probability, the average payoff will always
remain close to $E$. When both players observe their opponent's
moves (or at least the payoffs), Blackwell~\cite{BlackwellAnalogue}
proved that if $E$ satisfies some geometrical condition -- $E$ is then called a $B$-set --, then Player~1 can
approach it. He also deduced that  either Player~1 can approach a
convex set  or Player~2 can exclude it, i.e.\ he can approach the
complement of one of its neighborhood.

In the partial monitoring case,  when players do not observe
their opponent's moves but receive random signals (their laws may
depend on the actions played), working on the space of unknown payoffs might
not be sufficient -- except for specific cases, such as the minimization of external regret as did Lugosi, Mannor and Stoltz \cite{LugosiMannorStoltz}.

Attempts were made to circumvent this issue, notably by Aumann and
Maschler~\cite{aumannmaschler} and Kohlberg~\cite{kohlberg} in the
framework of repeated games with incomplete information. Lehrer and
Solan~\cite{LehrerSolanPSE} also considered strategies that are
defined, not as a  function of the unknown past payoffs, but as a
function of the past signals, and they proved the existence of
strategies that satisfy an extension of the consistency  property.
Perchet~\cite{PerchetApproachMOR} also used this approach to provide
a complete characterization of approachable convex  sets; it
extends Blackwell's one in the full monitoring case.

\subsection*{Games with partial monitoring}

Formally, we consider a two player repeated game $\Gamma$ with partial monitoring where, at stage $n
\in \mathds{N}$, Player~1  chooses an action $i_n$
in a finite set $I$ and, simultaneously, Player 2 chooses $j_n \in J$. This generates a vector
payoff $\rho_n=\rho(i_n,j_n) \in \mathds{R}^k$ where $\rho$ is a
mapping from $I \times J$ to $\mathds{R}^k$, extended to
$\Delta(I)\times \Delta(J)$ by
$\rho(x,y)=\mathds{E}_{x,y}[\rho(i,j)]:=\sum_{i,j} x_iy_j\rho(i,j)$, where $\Delta(I)$ and
$\Delta(J)$ stand for the sets of probability measures over $I$ and
$J$.

The important difference with usual repeated games with full monitoring is that, at stage $n$, Player~1 does not observe Player~2's action $j_n$, nor his
payoff $\rho_n$, but he receives a random signal $s_n \in S$ (where $S$ is the finite set of signals) whose law is
$s(i_n,j_n) \in \Delta(S)$. The mapping  $s : I \times J \to \Delta(S)$, known from both players, is also extended to $\Delta(I)\times \Delta(J)$ by
$s(x,y)=\mathds{E}_{x,y}[s(i,j)] \in \Delta(S)$. On the other hand, Player~2 observes $i_n$, $j_n$ and $s_n$.

In this framework,  a strategy $\sigma$ of Player~1 is a mapping from the set of past finite
observations $H^1:=\bigcup_{n \in \mathds{N}}(I \times S)^n$ into
$\Delta(I)$; similarly, a strategy $\tau$ of Player 2 is a mapping from $H^2:=\bigcup_{n \in \mathds{N}}(I \times
S\times J)^n$ into $\Delta(J)$. As usual, a couple of strategies
$(\sigma,\tau)$ generates a probability, also denoted by
$\mathds{P}_{\sigma,\tau}$ on $H^{\infty}:=\left(I \times S \times
J\right)^{\infty}$ endowed with the cylinder topology.

We introduce the so-called {\em maximal informative} mapping $\mathbf{s}$ from $\Delta(J)$  to
$\Delta(S)^I$ by $\mathbf{s}(y)=\left(s(i,y)\right)_{i \in I}$. Its
range $\mathcal{S} \subset \Delta(S)^I$ is a polytope (i.e.\ the
convex hull of a finite number of points) and any of its element  is
called a {\em flag}. Whatever being his move, Player~1 cannot
distinguish between two actions $y_0,y_1 \in \Delta(J)$ that
generate the same flag $\mu \in \mathcal{S}$, i.e.\ such that
$\mathbf{s}(y_0)=\mathbf{s}(y_1)=\mu$, thus $\mathbf{s}(y)$
-- although  not observed -- is the maximal information available to
Player~1, given $y \in \Delta(J)$. Note that with full monitoring, a
flag is simply the law of the action of Player~2.

\subsection*{Approachability}

Given a closed set $E \subset \mathds{R}^k$ and $\delta\geq0$, we
denote by $d(z,E)=\inf_{e \in E}\|z-e\|$ (with $\|\cdot\|$ the
Euclidian norm) the distance to $E$, by $E^{\delta}=\{\omega \in
\mathds{R}^k;\ d(\omega,E) < \delta\}$ the $\delta$-neighborhood of
$E$ and finally by $\Pi_{E}(z) =\left\{e \in E;\ d(z,E)=\| z -e \|\right\}$
the set of closest points to $z$ in $E$ (called projections of $z$). Given any sequence $\left\{a_m\right\}_{m \in \mathds{N}}$ and $n \geq 1$, $\overline{a}_n=\frac{1}{n}\sum_{m=1}^n a_m$ is its average up to the $n$-th term.

Blackwell~\cite{BlackwellAnalogue} defined approachability as follows. A closed  set $E \subset \mathds{R}^k$ is approachable by Player~1 if for every  $\varepsilon >0$, there exist a strategy  $\sigma$ of Player~1 and $N \in \mathds{N}$, such that for every strategy  $\tau$ of Player~2:
\[ \sup_{n \geq N} \mathds{E}_{\sigma,\tau}\left[d(\overline{\rho}_n,E)\right]\leq \varepsilon .\]
In a dual way,, a set $E$ is excludable by Player~2, if there exists  $\delta>0$ such that the complement of  $E^{\delta}$ is approachable by Player~2.

In words, Player~1 can approach a set $E \subset \mathds{R}^k$ if he
has a strategy  such that  the average payoff  converges\footnote{The almost sure convergence can also be required, but to the cost of cumbersome notations.} to $E$, uniformly with respect to the strategies of
Player~2.

In the case of a convex set $C$, Blackwell~\cite{BlackwellAnalogue} and Perchet \cite{PerchetApproachMOR} (see also Kohlberg \cite{kohlberg} for a specific case) provided a complete characterization with, respectively, full and partial monitoring. Those results can be summarized thanks to the following notation (that will furthermore ease statements of generalized results). Let $X$ and $\Chi$ denote some \textit{informative actions spaces} and
$P$ be a multivalued application from $X \times \Chi$ to
$\mathds{R}^k$. In our case, $X=\Delta(I)$ and $\Chi=\cS$ and $P(x,\xi)=\left\{ \rho(x,y);\ y \in \mathbf{s}^{-1}(\xi)\right\}$ for every $x \in X$ and $\xi \in
\Chi$; with full monitoring, $\Chi$ reduces to $\Delta(J)$ and $P(x,y)=\{\rho(x,y)\}$.

Both conditions of Blackwell \cite{BlackwellAnalogue} and Perchet~\cite{PerchetApproachMOR} reduce to the following succinct one:
\begin{equation}\label{condperchet} \mbox{A convex set}\ C\ \mbox{ is approachable by Player~1 if and only if}\
 \forall\, \xi \in \Chi,\, \exists\, x \in X,\ P(x,\xi) \subset C.\end{equation}
Actually, and as we shall see later, this result holds when $X$ and $\Chi$ are any convex compact sets of some Euclidian spaces and $P: X \times \Chi \rightrightarrows \mathds{R}^k$ is a {\em $L$-Lipschitzian convex hull}. The latter condition means that for every $x \in X$, $P(x,\cdot)$ is convex and that there exists  a family $\{p_{\kappa}: X \times \Chi \to \mathds{R}^k;\ \kappa \in \mathcal{\mathcal{K}}\}$ of $L$-Lipschitzian functions such that, for every $(x,\xi) \in X \times \Chi$, $P(x,\xi)=\co \big\{p_{\kappa}(x,\xi);\ \kappa \in \mathcal{K}\big\}$,  where $\co(\cdot)$ stands for the convex
hull.

\subsection*{Purely informative game}\label{sectionpurelyinformativegame}

The introduction of two arbitrary compact sets $X$ and $\Chi$, endowed with the weak-$\star$ topology, and a multivalued mapping $P$ motivate the following definition of the abstract {\em purely informative game} $\widetilde{\Gamma}$. At stage $n \in \mathds{N}$, Player~1 chooses $\mathbf{x}_n \in \Delta(X)$,  the set of probability
measures over $X$ and, simultaneously, Player~2 chooses $\boldsymbol{\xi}_n \in \Delta(\Chi)$.  Those choices generate the
outcome (the term payoff  will only be used  in $\Gamma$):
\[\theta_n=\theta(\mathbf{x}_n,\boldsymbol{\xi}_n)=\mathbf{x}_n\otimes \boldsymbol{\xi}_n \in \Delta(X \times \Chi),\]
where $\otimes$ stands for the product distribution. A strategy $\sigma$ of Player~1 is now a mapping from $\bigcup_{n \in
\mathds{N}}\left(\Delta(X)\times \Delta(\Chi)\right)^n$ to
$\Delta(X)$ and similarly, a strategy $\tau$ of Player~2 is defined as
a mapping from $\bigcup_{n \in \mathds{N}}\left(\Delta(X)\times
\Delta(\Chi)\right)^n$ to $\Delta(\Chi)$. With these notations, a pair of strategies
$(\sigma,\tau)$ induces a unique sequence
$\left(\mathbf{x}_n,\boldsymbol{\xi}_n\right)_{n \in \mathds{N}}$ in
$\left(\Delta(X)\times \Delta(\Chi)\right)^{\mathds{N}}$.

Let $\overline{\theta}_n$  be the average up to stage $n$ of the measures $\theta _n$ which is defined as follows: for every Borel
subset $F \subset X \times \Chi$,
$\overline{\theta}_n(F):=\frac{1}{n}\sum_{m=1}^n \theta_m(F)$. Then a closed set $\widetilde{E} \subset \Delta(X \times \Chi)$ is
approachable by Player~1 if for every $\varepsilon >0$ there exist
a strategy $\sigma$ of Player~1 and $N \in \mathds{N}$ such that for
every strategy $\tau$ of Player~2:
\[\forall n \geq N, W_2\left(\overline{\theta}_n,\widetilde{E}\right):=\inf_{ \theta \in \widetilde{E}}W_2(\overline{\theta}_n,\theta) \leq \varepsilon,\]
where $W_2$ is the (quadratic) Wasserstein distance -- defined later in a section devoted to preliminaries  --. For the definition of approachability,   any distance  metrizing the weak-$\star$ convergence of measures could be suitable but for the characterization of approachability  - as we will demonstrate throughout  the paper - the quadratic Wasserstein distance is very convenient.

\subsection*{Organization and main results}  The purely informative game $\widetilde{\Gamma}$ unifies the framework of both games with or without signals; we prove indeed in the first section (see Proposition \ref{tildenormalequiv}) that a set is approachable in a game $\Gamma$   if and only if its {\em image set}  is also approachable in $\widetilde{\Gamma}$.
And we exhibit in the second section, see Proposition \ref{tildecond}, a necessary and sufficient condition under which the latter holds. So this gives immediately the same result for (non necessarily convex) approachable sets with partial monitoring, for the first time in the literature.

We investigate the case of approachable convex sets, which usually benefits from a quite simple characterization (condition  (\ref{condperchet})). We show in Section 3 that this is covered by our first main result, Theorem \ref{tildeconvexe} -- that is actually more general.

In the last section, we are interested in specific games (that we called convex) and thanks to a totally different notion of approachability (along with some unusual definition of convexity) we obtain, not only another characterization of approachable sets, but also rates of converges. Those main results are stated in Theorem  \ref{theobhatsetapproch} and Theorem \ref{theoconvexhatapproch}.

\section{Some preliminaries on Wasserstein Distance and on Normals}

Here we define in a precise and concise  way the distance $W_2$ we have already used in the introduction.
We also introduce some material that will be used in the sequel. The reader can refer for this part to the books \cite{Villani, Dudley}. Moreover  the projection onto a nonconvex set in $\R^k$ is in general multi valued and a proper notion of normal should be used (cf for instance \cite{AsSoulaimaniQuincampoixSorin}). So we need to adapt this definition to the set of measures (following ideas of \cite{CardaliaguetQuincampoix}).

 For every $\mu$ and $\nu$ in $\Delta^2\left(\mathds{R}^N\right)$ -- the
set of measures with a finite moment of order 2 in some euclidian
space $\mathds{R}^N$ --, the (squared) Wasserstein distance between $\mu$ and $\nu$ is defined by:
\begin{equation}\label{defwass}W^2_2(\mu,\nu):=\inf_{\gamma \in \Pi(\mu,\nu)}I[\gamma]=\sup_{(\phi,\psi) \in \Xi }J(\phi,\psi)=\inf_{U \sim \mu ; V \sim\nu}\mathds{E}\left[\|U-V\|^2\right], \ \mbox{where}\end{equation}
\begin{itemize}
\item[--]{ $U\sim\mu$ means that the law of the random variable $U$ is
$\mu$;}
\item[--]{ $\Pi(\mu,\nu)$ is the set of probability measures $\gamma \in \Delta\left(\mathds{R}^N \times \mathds{R}^N\right)$ with first marginal $\mu$ and second marginal $\nu$ and  \[I[\gamma]=\int_{\mathds{R}^N \times \mathds{R}^N}\|x-y\|^2\mathrm d\gamma(x,y)\,;\]}
\item[--]{$\Xi$ is the set of functions $(\phi,\psi) \in  L^1_\mu(\mathds{R}^N, \mathds{R})\times L^1_\nu(\mathds{R}^N, \mathds{R})$ such that  $\phi(x)+\psi(y) \leq \|x-y\|^2,\ \mu\otimes\nu$-as and \[ J(\phi,\psi)=\int_{\mathds{R}^N}\phi \mathrm d\mu + \int_{\mathds{R}^N} \psi \mathrm d\nu.\]
Furthermore, if the supports of $\mu$ and $\nu$ are included in a compact set, then we can assume that $\Xi$ is reduced to the set of  functions $(\phi,\phi^*)$ such that, for some arbitrarily fixed $x_0 \in K$,
\[ \phi^*(x)=\inf_{y \in K}\|x-y\|^2-\phi(y), \mathrm{ \ } \phi=\left(\phi^*\right)^* \mathrm{\ and \ } \phi(x_0)=0.
\]
Since every function in $\Xi$ is $2\|K\|$-lipschitzian (where $\|K\|$ is the diameter of $K$), Arzela-Ascoli's theorem implies that $\left(\Xi,\|\|_{\infty}\right)$ is relatively compact.}
\end{itemize}

Actually, the infimum and supremum in (\ref{defwass}) are achieved;  we denote by $\Phi(\mu,\nu)$ the subset of $\Xi$ that maximizes
$J(\phi,\phi^*)$ and  its elements are called {\em Kantorovitch potentials} from
$\mu$ to $\nu$. Any probability measure $\gamma \in
\Delta\left(\mathds{R}^{2N}\right)$ that achieves the minimum  is
 an  {\em optimal plan} from $\mu$ to $\nu$.

More details on the definition of $W_2$, based on Kantorovitch duality, can be found for example in Dudley~\cite{Dudley}, chapter~11.8 or Villani~\cite{Villani},
chapter~2.

Brenier's~\cite{Brenier} theorem states that if $\mu \ll_0 \boldsymbol{\lambda}$ (i.e., the probability measure $\mu$ is absolutely continuous with respect to the Lebesgue measure $\boldsymbol{\lambda}$ and has a strictly positive density), then there exist a unique optimal plan
$\gamma$ and a unique convex Kantorovitch potential from $\mu$ to $\nu$. They  satisfy:
\[d\gamma(x,y)=d\mu(x)\delta_{\left\{x-\nabla\phi(x)\right\}} \mbox{\ or \ equivalently\ } \gamma=\left(\Id \times (\Id-\nabla\phi)\right) \sharp \mu,\]
where for any $\psi:
\mathds{R}^N \to  \mathds{R} ^N$, Borel measurable with at most a
linear growth, $\psi\sharp \mu \in
\Delta\left(\mathds{R}^N\right)$ is the push-forward of $\mu$ by
$\psi$ --  also called the image probability measure of $\mu$
by $\psi$. It is defined by
 \[ \psi\sharp \mu(A)= \mu\left(\psi^{-1}(A)\right)\qquad \forall A
 \subset \mathds{R}^N, \; \mathrm{Borel\ measurable} \]
 or equivalently by: for every Borel measurable bounded maps  $f: \mathds{R}^N\to \mathds{R}$:
 \[\int_{\mathds{R}^N} f\mathrm d (\psi\sharp \mu)= \int_{\mathds{R}^N} f(\psi(x))\mathrm d\mu(x).\]

A classical approximation result (see e.g.\ Dudley~\cite{Dudley})) is that for any convex compact $K \subset \R^N$ with non-empty interior and any $\varepsilon>0$, there exists
$\Delta_{\varepsilon}(K)$ a compact subset of $\Delta_0(K)=\left\{\mu \in
\Delta(K), \mu \ll_0 \boldsymbol{\lambda}\right\}$ such that, for every $\mu \in \Delta(K)$, $W^2_2(\mu,\Delta_{\varepsilon}(K))\leq \varepsilon$. So Brenier's theorem actually implies that
\[\Phi\ \mbox{is singlevalued and uniformly continuous on}\ \Delta_{\varepsilon}(K)\times K.
\]

\subsection*{Some geometrical properties of $W_2$} Blackwell's approachability results rely deeply on the geometry of Euclidian spaces. One of our goals is to underline and establish required results for the measure space equipped with $W_2$. For instance, in Euclidian (and also Hilbert) spaces the projection onto a closed convex set could be characterized equivalently by the minimization of the distance to the set or by a characterization of the projection by the well-known condition with scalar products.   The Lemma \ref{convex-Cond} could be viewed as a way of writing this  "condition with scalar products" in the space of measures. Also Blackwell's conditions requires suitable notions of projections and  normals we will introduce now.

   In the space $\Delta(K)$ equipped with $W_2$ we define  the corresponding definitions of proximal normals (see  Bony~\cite{Bony})  to  nonempty closed sets $A \subset \Delta(K)$ at some $\underline{\mu} \in A$. As usual, we say that  $\underline{\mu}$ is a projection (with respect  to the Wasserstein distance) of a measure $\mu \in \Delta(K)$ if $W_2 ( \mu ,A):= \inf_{\theta \in A}W_2(\mu,\theta)=W_2 ( \mu,\underline{\mu})$.

Actually, proximal normals  can be defined in two different ways, depending on
which equivalent  definition of $W_2$ is used.

\begin{itemize}
\item[--]{{\em Proximal potential normal}: a continuous  function $\phi : K \to \mathds{R}$ is a
proximal potential normal to $A$ at $\underline{\mu}$ if $\phi$ is a Kantorovitch potential
from $\underline{\mu}$ to some $\mu \in \Delta(K)$ where $\underline{\mu}$ is a
projection of $\mu$ on $A$.
\[ \NC^p_{A}\left(\underline{\mu}\right) = \left\{ \phi: K \to \R;\ \phi\ \mbox{proximal potential
normals to}\ A\ \mbox{at}\ \underline{\mu}\right\}.\] }
\item[--]{ {\em Proximal gradient normal} (adapted from \cite{Soulaimani}): a map $p\in L_{\underline{\mu}}^{2}(\mathds{R}^{N},\mathds{R}^{N})$ is a
proximal gradient normal to $A$ at $\underline{\mu}$ if  $p \in
\mathcal{P}^\gamma(\underline{\mu},\mu)$ for some  $\mu \not\in A$ that projects on $\underline{\mu}$ and  some optimal plan $\gamma \in \Pi(\underline{\mu},\mu)$, where, for every $\mu, \nu  \in \Delta(\R^N)$ and $\gamma \in \Pi(\mu,\nu)$,
\[\mathcal{P}^\gamma(\underline{\mu},\mu)=\left\{ p\in L^2_\mu\left(\mathds{R}^N,\mathds{R}^N\right)\,;\ \forall \psi \in \mathcal{C}_{LB},  \int_{\mathds{R}^N} \langle \psi(x), p(x)\rangle\mathrm d\mu(x)=\int_{\mathds{R}^{2N}}
 \langle\psi(x),x-y\rangle\mathrm d\gamma(x,y)\right\}\] with $\mathcal{C}_LB$ the sets of  Borel measurable map $\psi:\mathds{R}^N\to\mathds{R}^N$ with at most a
 linear growth. Riesz representation theorem ensures the non-emptiness of $\mathcal{P}^\gamma(\underline{\mu},\mu)$ (see \cite{CardaliaguetQuincampoix}).
\[ \NC^g_{A}\left(\underline{\mu}\right) = \left\{ p \in  L_{\underline{\mu}}^{2}(\mathds{R}^{N},\mathds{R}^{N});\ p\ \mbox{proximal gradient normals to}\ A\ \mbox{at}\ \underline{\mu}\right\}.\]}
\end{itemize}

Observe that Brenier's Theorem also implies that both
definitions of proximal normals are, in some sense, quite close. Indeed, if $A$ is a compact subset of $\Delta(K)$ and $\underline{\mu} \in
A$ and $\mu \ll_0 \boldsymbol{\lambda}$   then
\[\phi \in \NC_A^p(\underline{\mu}) \Longrightarrow \nabla \phi \in \NC_A^g(\underline{\mu}).\]

\section{Equivalences between approachability in both games}\label{sectionapproachabilityPurely}

Recall that we represent a game $\Gamma$ by two compact convex action spaces $X$ and $\Chi$ and a $L$-Lipschitzian convex hull $P$. We define the  set of  outcomes compatible  with $\theta \in
\Delta(X \times \Chi)$ by:
\[\rho(\theta)=\int_{X \times \Chi} P(x,\xi)\mathrm d\theta \subset \mathds{R}^k\]
where the integral is in Aumann~\cite{AubinFrankowska}'s sense: it is the set of all  integrals of
measurable selection of $P$. For every subset $\widetilde{E} \subset
\Delta(X \times \Chi)$, the set of compatible outcomes
$\rho(\widetilde{E})\subset \mathds{R}^k$ is defined by:
\[\rho(\widetilde{E})=\left\{\rho(\theta);\ \theta \in \widetilde{E}\right\} \subset \mathds{R}^k.\]
 Reciprocally, for every $E \subset \mathds{R}^k$, the set of compatible measures $\rho^{-1}(E)\subset \Delta(X \times \Chi)$  is defined by:
\[\rho^{-1}(E)=\left\{\theta \in \Delta(X \times \Chi);\ \rho(\theta) \subset E\right\}.\]

The mapping $\mathbf{s}$ does not appear in the description
of $\widetilde{\Gamma}$ but only in the definition of
$\rho$ that links  $\Gamma$ and
$\widetilde{\Gamma}$: given a set $E \subset \mathds{R}^k$ in
$\Gamma$, we introduced the image set  $\widetilde{E}=\rho^{-1}(E) \subset \Delta(X \times
\Chi)$ in $\widetilde{\Gamma}$. And it is quite intuitive that $E$ is
approachable if and only $\widetilde{E}$ is (see Proposition
\ref{tildenormalequiv} below, whose proof -- mainly technical -- is delayed to the Appendix in order to keep some fluency).

 \begin{proposition}\label{tildenormalequiv}\
\begin{enumerate}
\item[i)]{A set $E \subset \mathds{R}^k$ is approachable in $\Gamma$ if and only if  $\rho^{-1}(E) \subset \Delta\left(X\times\Chi\right)$ is approachable in $\widetilde{\Gamma}$;}
\item[ii)]{If a set $\widetilde{E} \subset \Delta\left(X\times\Chi\right)$ is approachable in $\widetilde{\Gamma}$ then the set $\rho(\widetilde{E}) \subset \mathds{R}^k$ is approachable in $\Gamma$;}
\item[iii)]{For every convex set $C \subset \mathds{R}^k$, $\rho^{-1}(C) \subset \Delta\left(X\times\Chi\right)$ is a (possibly empty) convex set and for every convex set $\widetilde{C} \subset \Delta\left(X\times\Chi\right)$, $\rho(\widetilde{C})$ is a convex set.}
    \end{enumerate}
\end{proposition}

Notice that point $ii)$ cannot be an equivalence. Consider  the case $\rho=0$ and $\widetilde{E}=\left\{\mathbf{x_0}\otimes
\mathbf{y_0}\right\}$ for some $\mathbf{x_0}$ and $\mathbf{y_0}$. Then  $\rho(\widetilde{E})=\{0\}$ is
approachable but $\widetilde{E}$ is not, since Player~2 just has to
play $\mathbf{y_1} \neq \mathbf{y_0}$ at each stage. This is a consequence of the usual inclusions $\rho\left(\rho^{-1}(E)\right) \subset E$ and $\rho^{-1}\left(\rho(\widetilde{E})\right) \supset \widetilde{E}.$

\section{Approachability of $\widetilde{B}$-sets}\label{sectioncharacPurely}
Blackwell~\cite{BlackwellAnalogue} noticed that a
closed set $E$ that fulfills the following geometrical condition  -- $E$ is then call a $B$-set -- is approachable by
Player  1 with full monitoring.  Formally, a closed subset  $E$  of $\mathds{R}^k$ is a  $B$-set, if \[ \forall z \in \mathds{R}^k,\, \exists p \in \Pi_{E}(z),\,  \exists x \in \Delta(I),\   \langle \rho(x,y) - p, z - p\rangle \leq 0, \quad \forall y \in \Delta(J).\]
An equivalent formulation using $\NC_{E}(q)$, the set of proximal normals to $E$ at $q$, appeared in \cite{AsSoulaimaniQuincampoixSorin}. Indeed $E$ is a $B$-set if and
only if
\[\forall p \in E,\, \forall q \in \NC_E(p),\, \exists x \in \Delta(I),\, \forall y \in \Delta(J),\ \langle \rho(x,y)-p,q\rangle \leq 0.\]
Blackwell \cite{BlackwellAnalogue} and  Spinat \cite{Spinat} proved that $E$ is approachable in $\Gamma^f$ if and only if it contains a
$B$-set.

Our definition of proximal potential normals gives to $\mathbf{W}_2$ a structure close to a Hilbert. This allows to extend Blackwell's definition of a $B$-set as follows.

\begin{definition}
A set $\widetilde{E} \subset \Delta(X \times \Chi)$ is a
$\widetilde{B}$-set if for every $\theta$ not in
$\widetilde{E}$ there exist $\underline{\theta} \in
\Pi_{\widetilde{E}}(\theta)$, $\phi \in
\Phi(\underline{\theta},\theta)$ and
$\mathbf{x}\left(=\mathbf{x}(\theta)\right) \in
\Delta(X)$ such that :
\[ \int_{X \times \Chi} \phi  \ \mathrm d(\underline{\theta}- \mathbf{x}\otimes\boldsymbol{\xi}) \leq 0, \quad \forall \boldsymbol{\xi} \in \Delta(\Chi).\]
Or stated in terms of proximal potential normals:
\[\forall \underline{\theta} \in \widetilde{E},\, \forall \phi \in \NC_{\widetilde{E}}^p(\underline{\theta}),\, \exists \mathbf{x} \in \Delta(X),\, \forall \boldsymbol{\xi} \in \Delta(\Chi),\ \int_{X \times \Chi} \phi  \ \mathrm d(\underline{\theta}- \mathbf{x}\otimes\boldsymbol{\xi}) \leq 0.\]
\end{definition}

The concept of $\widetilde{B}$-sets is indeed the natural extension of
 $B$-sets because of the following proposition.

\begin{proposition}\label{tildecond}
A set $\widetilde{E} \subset \Delta\left(X\times\Chi\right)$ is
approachable if and only if it contains a $\widetilde{B}$-set.
\end{proposition}
\textbf{Proof:} We only prove here the sufficient part, i.e.\ a
$\widetilde{B}$-set is approachable by Player~1 (by adapting
Blackwell~\cite{BlackwellAnalogue}'s ideas to our framework). Again, we postpone the proof of the necessary part (almost identical to the full monitoring case) to the Appendix to prevent cumbersomeness.

\medskip

Let $\varepsilon >0$ be fixed. For every probability distribution
$\theta \in \Delta(X \times \Chi)$, we denote by $\theta^{\varepsilon} \in \Delta_{\varepsilon}(X\times \Chi)$ any arbitrary
approximation of $\theta$  such that
$W_2^2(\theta,\theta^{\varepsilon})\leq \varepsilon$.

Consider the strategy $\sigma^{\varepsilon}$ of Player~1 that plays, at stage $n \in
\mathds{N}$, $\mathbf{x}(\bar{\theta}_{n-1}^{\varepsilon})$ given by the
definition of a $\widetilde{B}$-set, where $\bar{\theta}_{n-1}^{\varepsilon}$ is the average of the $n-1$ first $\left(\theta_{m}\right)^{\varepsilon}$. Then, if we denote  by
$\underline{\theta}_{n-1}^{\varepsilon}$ the projection of
$\bar{\theta}_{n-1}^{\varepsilon}$ over $\widetilde{E}$ and let
$w_n=W_2^2\left(\bar{\theta}_n^{\varepsilon},\widetilde{E}\right)$:
\vspace{-0.5cm}
\begin{align*}
w_n=W_2^2\left(\bar{\theta}_n^{\varepsilon}, \widetilde{E}\right)&\leq W_2^2\left(\bar{\theta}_n^{\varepsilon},\underline{\theta}_{n-1}^{\varepsilon}\right)=W_2\left(\underline{\theta}_{n-1}^{\varepsilon},\frac{n-1}{n}\bar{\theta}^{\varepsilon}_{n-1}+\frac{\theta^{\varepsilon}_{n}}{n}\right)\\
&= \sup_{\phi \in \Xi}\int_{X \times \Chi} \phi\ \mathrm d\underline{\theta}_{n-1}^{\varepsilon}+\int_{X \times \Chi}\phi^* \mathrm d\left(\frac{n-1}{n}\bar{\theta}^{\varepsilon}_{n-1}+\frac{\theta^{\varepsilon}_{n}}{n}\right)\\
&=\int_{X \times \Chi} \phi_n\ \mathrm d\underline{\theta}_{n-1}^{\varepsilon}+\int_{X \times \Chi}\phi_n^* \mathrm d\left(\frac{n-1}{n}\bar{\theta}^{\varepsilon}_{n-1}+\frac{\theta^{\varepsilon}_{n}}{n}\right)\\
&\leq \frac{n-1}{n} w_{n-1}+ \frac{1}{n}\left( \int_{X \times \Chi}
\phi_n \mathrm d\underline{\theta}_{n-1}^{\varepsilon} - \int_{X
\times \Chi}\phi_n^* \mathrm d \theta^{\varepsilon}_{n}\right)
\end{align*}
where $\phi_{n}$ is the optimal Kantorovitch potential from
$\underline{\theta}_{n-1}^{\varepsilon}$ to
$\frac{n-1}{n}\bar{\theta}^{\varepsilon}_{n-1}+\frac{\theta^{\varepsilon}_{n}}{n}$.
Let us denote by $\phi_0$ the optimal Kantorovitch potential from
$\underline{\theta}_{n-1}^{\varepsilon}$ to
$\bar{\theta}^{\varepsilon}_{n-1}$ and by
$\omega^{\varepsilon}(\cdot)$ the modulus of continuity of $\Phi$
restricted to the compact set $ \widetilde{E} \times
\Delta_{\varepsilon}(X \times \Chi)$.

The definition of $W^2_2$ implies that
\[W^2_2\left(\bar{\theta}^{\varepsilon}_{n-1},\frac{n-1}{n}\bar{\theta}^{\varepsilon}_{n-1}+\frac{\theta^{\varepsilon}_{n}}{n}\right)
\leq
\frac{1}{n}W_2^2(\bar{\theta}^{\varepsilon}_{n-1},\theta^{\varepsilon}_{n})\leq
\frac{4\|X\times\Chi\|^2}{n},\] therefore  $\|\phi_0-\phi_n\|_{\infty} \leq
\omega^{\varepsilon}(2\|X\|/\sqrt{n})$ and
\[w_{n}\leq \frac{n-1}{n}w_{n-1}+\frac{1}{n}\left(\int_{X \times \Chi}\phi_{0} \mathrm d\underline{\theta}_{n-1}^{\varepsilon}+\int_{X\times \Chi} \phi_{0}^*\mathrm d\theta^{\varepsilon}_{n}\right)+\frac{2}{n}\omega\left(\frac{2\|X\times\Chi\|}{\sqrt{n}}\right).\]
Recall that $\theta_{n}^{\varepsilon}$ is such that that $\int_{X
\times \Chi} \phi_0\mathrm d\theta_{n}+\int_{X \times \Chi}
\phi_0^*\mathrm d\theta_{n}^{\varepsilon} \leq
W^2_2(\theta_{n},\theta_{n}^{\varepsilon},)\leq \varepsilon $,
therefore
\[w_{n}\leq \frac{n-1}{n}w_{n-1}+\frac{1}{n}\left(\int_{X \times \Chi}\phi_{0} \mathrm d\underline{\theta}_{n-1}^{\varepsilon}-\int_{X\times \Chi} \phi_{0}\mathrm d\theta_{n} \right)+\frac{2}{n}\omega\left(\frac{2\|X\times\Chi\|}{\sqrt{n}}\right)+\frac{\varepsilon}{n}.\]
Since $\widetilde{E}$ is a $\widetilde{B}$-set and because of the
choice of $\mathbf{x}_n \in \Delta(X)$,  for every
$\boldsymbol{\xi}_n \in \Delta(\Chi)$,  $\int_{X\times
\Chi}\phi_0\mathrm d\left(\underline{\theta}_{n-1}^{\varepsilon}-
\mathbf{x}_n\otimes\boldsymbol{\xi}_n\right) \leq 0$, thus
\[w_{n} \leq \frac{n-1}{n}w_{n-1}+\frac{2}{n}\omega^{\varepsilon}\left(\frac{2\|X\times\Chi\|}{ \sqrt{n}}\right)+\frac{\varepsilon}{n}\]
and  this yields, by induction, that
\[W_2^2\left(\bar{\theta}_n^{\varepsilon}, \widetilde{E}\right) \leq \frac{1}{n+1}W^2_2\left(\overline{\theta^{\varepsilon}}_{1}, \widetilde{E}\right)+\frac{2}{n+1}\sum_{m=1}^{n}\omega^{\varepsilon}\left(\frac{2\|X\times\Chi\|}{\sqrt{k}}\right)+\varepsilon.\]
Since
$\omega^{\varepsilon}(2\|X\times\Chi\|/\sqrt{k})$ converges to 0 when $k$ goes
to infinity, then $w_n$ is asymptotically smaller than $\varepsilon$.  The fact that $W_2\left(\overline{\theta}_{n},
\widetilde{E}\right)\leq W_2\left(\bar{\theta}_n^{\varepsilon},
\widetilde{E}\right)+\varepsilon$ implies that  $\widetilde{E}$ is
approachable by Player~1.
  $\hfill \Box$

\section{Characterization of convex approachable sets}\label{sectionCharacConvexPurely}
 There also exists in $\widetilde{\Gamma}$ a complete characterization of approachable convex sets :
\begin{theorem}\label{tildeconvexe}
A convex  set $\widetilde{C}$ is approachable if and only if:
\[ \forall \boldsymbol{\xi} \in \Delta(\Chi),\, \exists \mathbf{x}\in \Delta(X),\  \theta(\mathbf{x}, \boldsymbol{\xi})=\mathbf{x}\otimes\boldsymbol{\xi} \in \widetilde{C}.\]
\end{theorem}
\textbf{Proof:} Once again, we will follow the idea of Blackwell.
Assume that there exists $\boldsymbol{\xi}$ such that, for every
$\mathbf{x} \in \Delta(X)$, $\theta(\mathbf{x},\boldsymbol{\xi})
\notin \widetilde{C}$. The application $\mathbf{x} \mapsto
W_2\left(\theta(\mathbf{x},\boldsymbol{\xi}),\widetilde{C}\right)$ is continuous on the
compact set $\Delta(X)$, therefore there exists $\delta>0$ such that
$W_2^2\left(\theta(\mathbf{x},\boldsymbol{\xi}),\widetilde{C}\right)
\geq \delta$.

Consider the strategy of Player~2 that consists of  playing
$\boldsymbol{\xi}$  at every stages, then
$\theta_n=\mathbf{x}_n\otimes \boldsymbol{\xi}$,
$\overline{\theta}_n=\overline{\mathbf{x}}_n\otimes
\boldsymbol{\xi}=\theta(\overline{\mathbf{x}}_n,\boldsymbol{\xi})$
and $W_2^2\left(\overline{\theta}_n,\widetilde{C}\right)\geq
\delta>0$. Therefore $\widetilde{C}$ is not approachable by Player
1.

\medskip

Reciprocally, assume that for every $\boldsymbol{\xi} \in
\Delta(\Chi)$ there exists $\mathbf{x} \in \Delta(X)$ such that
$\theta(\mathbf{x},\boldsymbol{\xi}) \in \widetilde{C}$. We claim
that this implies that $\widetilde{C}$ is a $\widetilde{B}$-set.

Let $\overline{\theta}$ be a probability measure that does not
belong to $\widetilde{C}$ and assume (for the moment) that $\overline{\theta} \ll_0 \boldsymbol{\lambda}$. Denote
by $\underline{\theta} \in \widetilde{C}$ any of its projection
then, by definition of the projection and convexity of
$\widetilde{C}$:
\begin{align*}W_2(\underline{\theta},\overline{\theta})\leq W_2^2\left((1-\lambda)\underline{\theta}+\lambda \mathbf{x}\otimes \boldsymbol{\xi},\overline{\theta}\right)&=\sup_{\phi \in \Xi}\int_{X \times \Chi} \phi\mathrm d\left((1-\lambda)\underline{\theta}+\lambda\mathbf{x}\otimes\boldsymbol{\xi}\right)+\int_{X \times \Chi} \phi^* \mathrm d\overline{\theta}\\
&=\sup_{\phi \in \Xi}\int_{X \times \Chi}\phi\mathrm d\underline{\theta}+\int_{X \times \Chi} \phi^* \mathrm d\overline{\theta}-\lambda \int_{X \times \Chi} \phi \mathrm d(\underline{\theta}-\mathbf{x}\otimes \boldsymbol{\xi})\\
&=\int_{X \times \Chi} \phi_{\lambda} \mathrm d\underline{\theta}+\int_{X \times \Chi}\phi_{\lambda}^*\mathrm d\overline{\theta}-\lambda \int_{X \times \Chi} \phi_{\lambda} \mathrm d(\underline{\theta}-\mathbf{x}\otimes \boldsymbol{\xi})\\
&\leq \int_{X \times \Chi} \phi_{0} \mathrm d\underline{\theta}+\int_{X \times \Chi}\phi_0^*\mathrm d\overline{\theta}-\lambda \int_{X \times \Chi} \phi_{\lambda} \mathrm d(\underline{\theta}-\mathbf{x}\otimes \boldsymbol{\xi})\\
&=W_2^2(\overline{\theta},\underline{\theta})-\lambda \int_{X \times
\Chi} \phi_{\lambda} \mathrm d(\underline{\theta}-\mathbf{x}\otimes
\boldsymbol{\xi})
\end{align*}
where $\phi_{\lambda}$ (resp.\ $\phi_0$) is the unique potential
from $(1-\lambda)\underline{\theta}+\lambda
\mathbf{x}\otimes\boldsymbol{\xi}$ (resp.\ $\underline{\theta}$) to
$\overline{\theta}$. Therefore, for every $\lambda>0$, $\lambda
\int_{X \times \Chi} \phi_{\lambda} \mathrm
d(\underline{\theta}-\mathbf{x}\otimes \boldsymbol{\xi})\leq 0$.
Dividing by $\lambda>0$ yields:
\[\int_{X \times \Chi} \phi_{\lambda} \mathrm d(\underline{\theta}-\mathbf{x}\otimes \boldsymbol{\xi})\leq 0, \quad \forall \lambda >0.\]
Since $(1-\lambda)\underline{\theta}+\lambda \mathbf{x} \otimes
\boldsymbol{\xi}$ converges to $\underline{\theta}$, any
accumulation point of $(\phi_{\lambda})_{\lambda>0}$ has to belong
(for every $\mathbf{x}$ and $\boldsymbol{\xi}$) to
$\Phi(\underline{\theta},\overline{\theta})=\{\phi_0\}$.   Stated
differently, given $\phi_0 \in
\Phi(\underline{\theta},\overline{\theta})$, one has:
\[\max_{\boldsymbol{\xi} \in \Delta(\Chi)} \min_{\mathbf{x} \in \Delta(X)} g_{\phi_0}(\mathbf{x},\boldsymbol{\xi}):=\max_{\boldsymbol{\xi} \in \Delta(\Chi)} \min_{\mathbf{x} \in \Delta(X)}\int_{X \times \Chi} \phi_{0} \mathrm d(\underline{\theta}-\mathbf{x}\otimes \boldsymbol{\xi})\leq 0.\]
The function $g_{\phi_0}$ is linear in both of its variable, so
Sion's Theorem implies that \[\max_{\boldsymbol{\xi} \in
\Delta(\Chi)} \min_{\mathbf{x} \in \Delta(X)}
g_{\phi_0}(\mathbf{x},\boldsymbol{\xi})=  \min_{\mathbf{x} \in
\Delta(X)} \max_{\boldsymbol{\xi} \in
\Delta(\Chi)}g_{\phi_0}(\mathbf{x},\boldsymbol{\xi})\] hence
$\widetilde{C}$ is a $\widetilde{B}$-set.

Assume now that $\overline{\theta} \not \in \Delta_0(X\times\Chi)$ and let
$\overline{\theta}_n \in \Delta_{\frac{1}{n}}(X\times\Chi)$ be a sequence of
measures that converges to $\overline{\theta}$,
$\left(\underline{\theta}_n\right)_{n \in \mathds{N}}$ a sequence of their
projections, and $\phi_n \in
\Phi(\underline{\theta}_n,\overline{\theta}_n)$.  Up to subsequences, we can assume that $\underline{\theta}_n$ and $\phi_n$
converge respectively to $\underline{\theta}$ and $\phi_0$.
Necessarily, $\underline{\theta}$ is a projection of
$\overline{\theta}$ onto $\widetilde{C}$ and $\phi_0$ belongs to
$\Phi(\underline{\theta},\overline{\theta})$. Therefore:
\[\int_{X \times \Chi} \phi_0 \mathrm d(\underline{\theta}-\mathbf{x}\otimes \boldsymbol{\xi})=\lim_{n \to \infty} \int_{X \times \Chi} \phi_{n} \mathrm d(\underline{\theta}-\mathbf{x}\otimes \boldsymbol{\xi}) \leq 0\]
and $\widetilde{C}$ is a $\widetilde{B}$-set.  $\hfill \Box$

\bigskip

Let us go back and quickly show that the characterization (\ref{condperchet}) of  approachable convex sets with full monitoring is a
consequence of Theorem \ref{tildeconvexe}:

\textbf{Proof of characterization (\ref{condperchet})} By Proposition
\ref{tildenormalequiv}, a convex subset $C$
of $\mathds{R}^k$ is approachable if and only if the convex set
$\rho^{-1}(C) \subset \Delta\left(X\times\Chi\right)$ is
approachable in $\widetilde{\Gamma}$. Therefore, using Theorem
\ref{tildeconvexe}, if and only if for every $\boldsymbol{\xi} \in
\Delta(\Chi)$, there exists $\mathbf{x} \in \Delta(X)$ such that
$\rho(\mathbf{x}\otimes\boldsymbol{\xi}) \subset C$. Let us denote by $\mathds{E}_{\mathbf{x}} \in X$ and $\mathds{E}_{\boldsymbol{\xi}} \in \Chi$ the expectations of any $\mathbf{x} \in \Delta(X)$ and $\boldsymbol{\xi} \in \Delta(\Chi)$. Then, in the case of games with partial monitoring, one has that  $\rho(\delta_{\mathds{E}_\mathbf{x}}\otimes \delta_\xi)\subset \rho(\mathbf{x}\otimes \delta_\xi)$ and  $\rho(\delta_{x}\otimes\boldsymbol{\xi})\subset \rho(\delta_{x}\otimes\delta_{\mathds{E}_{\boldsymbol{\xi}} })$.

Assume that $C$ is approachable; since the condition holds in particular for $\boldsymbol{\xi}=\delta_\xi$, there exists $\mathbf{x}$ such that $\rho(\mathbf{x}\otimes \delta_\xi) \subset C$, therefore some $x \in X$ such that $\rho(\delta_x\otimes \delta_\xi) \subset C$ (one just  has to take $x=\mathds{E}_\mathbf{x}$).

Reciprocally, if $C$ is not approachable; there exists  $\boldsymbol{\xi}$ such that (in particular) for any $\delta_x$, $\rho(\boldsymbol{\xi}\otimes \delta_x) \not\subset C$,
therefore there exists some $\xi \in \Chi$ such that $\rho(\delta_\xi\otimes \delta_x) \not\subset C$ (one just has to take $\xi=\mathds{E}_{\boldsymbol{\xi}}$).

We obtain the stated result as a consequence:
\[ C \ \mbox{is approachable if and only if} \ \forall\, \xi \in \Chi,\, \exists\, x \in X,\ \rho(\delta_x\otimes\delta_\xi)=P(x,\xi) \subset C \]
$\hfill \Box$  

This also  explains why there exist convex sets that are
neither approachable, nor excludable with partial monitoring  (see
Perchet~\cite{PerchetApproachMOR}) which cannot occur with full monitoring
(see Blackwell~\cite{BlackwellAnalogue}): it simply due to the fact
that $\rho^{-1}(C)$ can be empty.

\section{Convex games}\label{convexgames}
We restrict ourselves in this section to the particular class of
games called  {\em convex games} which have the following
property:  for every $q \in \Delta(X \times \Chi)$:
\[ \rho(q)=\int_{X \times \Chi} P(x,\xi) \mathrm dq(x,\xi) \subset P\left(\mathds{E}_{q}\left[x\right],\mathds{E}_q\left[\xi\right]\right).\] For instance,  this reduces in games with full monitoring to $\rho(q)=\rho\left(\mathds{E}_{q}[x],\mathds{E}_{q}[\xi]\right)$.

\begin{example} The following game where the  payoffs of Player~1 are given by
the matrix on the left and signals by the matrix on the right, is
convex.
\begin{center}
\begin{tabular}{c|c|c|c|p{2cm}c |c|c|c|}
\multicolumn{1}{c}{}&\multicolumn{1}{c}{$L$}&\multicolumn{1}{c}{$C$}&\multicolumn{1}{c}{$R$}&\multicolumn{2}{c}{}&\multicolumn{1}{c}{$L$}&\multicolumn{1}{c}{$C$}&\multicolumn{1}{c}{$R$}\\
\cline{2-4} \cline{7-9}
$T$& (0,-1) & (1,-2)&(2,-4)&&$T$& $a$ & $a$ & $b$\\
\cline{2-4} \cline{7-9}
$B$& (1,0)& (2,-1) & (3,-3) & &$B$& $a$ & $a$& $b$\\
\cline{2-4} \cline{7-9}
\end{tabular}
\end{center}
In this game $I=\{T,B\}$, $J=\{L,C,R\}$ and $S=\{a,b\}$. If Player~1
receives the signal $a$, he does not know whether Player~2 used the
action $L$ or $C$. \end{example}
\bigskip

We  introduce the notions of displacement interpolation and
convexity (see e.g.  Villani~\cite{Villani} for more details)  that will play the role of classic
linear interpolation and convexity.

Given $\mu$, $\nu \in \Delta^2(\mathds{R}^N)$ and $t \in [0,1]$, a
displacement interpolation between $\mu$ and $\nu$ at time $t$ is
defined by $\widehat{\mu}_t= \sigma_{t}\sharp \gamma$,  where
$\gamma \in \Pi(\mu,\nu)$ is an optimal plan and $\sigma_t$ is the mapping defined by $
\sigma_t(x,y)=(1-t)x+ty$. A set $\widehat{C}$ is displacement convex if for every $\mu,\nu \in
\widehat{C}$, every $t \in [0,1]$ and every optimal plan $\gamma \in
\Pi(\mu,\nu)$, $\sigma_t\sharp \gamma \in \widehat{C}$.

Let  $\widehat{\Gamma}$ be a new game defined as follows. At stage $n \in
\mathds{N}$, Player~1 (resp.\ Player~2) chooses $x_n \in X$ (resp.\
$y_n \in \Chi$) and the payoff is
$\theta_n=\delta_{x_n}\otimes\delta_{y_n}=\delta_{(x_n,y_n)} \in
\Delta(X \times \Chi)$. We do not consider average payoffs in the
usual sense (as in $\widetilde{\Gamma}$) but we define  a sequence
of recursive interpolation by:  \[\widehat{\theta}_{n+1}=
\sigma_{\frac{1}{n+1}} \sharp\gamma_{n+1}, \mathrm{\ where \ }
\gamma_{n+1} \in \Pi(\widehat{\theta}_n,\theta_{n+1}) \mathrm{\ is \
an \ optimal \ plan.}\] By induction, this implies that
$\widehat{\theta}_n=\delta_{\overline{x}_n}\otimes\delta_{\overline{y}_n}$.
Indeed, $\widehat{\theta}_1=\delta_{x_1}\otimes\delta_{y_1}$ and
$\theta_2=\delta_{x_2}\otimes\delta_{y_2}$
therefore:\[\gamma_2=\left(\delta_{x_1}\otimes\delta_{y_1}\right)\otimes
\left( \delta_{x_1}\otimes\delta_{y_1}\right) \mathrm{\ and \ }
\widehat{\theta}_2=\sigma_{\frac{1}{2}}\sharp\gamma_2=\delta_{\frac{x_1+x_2}{2}}\otimes\delta_{\frac{y_1+y_2}{2}}.\]
\begin{definition}
A closed set $\widehat{E} \subset \Delta(X \times \Chi)$ is
displacement approachable by Player~1 if for every $\varepsilon >0$
there exist a strategy $\sigma$ of Player~1 and $N \in \mathds{N}$
such that for every strategy $\tau$ of Player~2:
\[\forall n \geq N, W_2\left(\widehat{\theta}_n,\widehat{E}\right) \leq \varepsilon.\]
\end{definition}

Consider any set $E \subset  \mathds{R}^d$ and assume that
$\rho^{-1}(E)$ is displacement approachable by Player~1.
 Since $\widehat{\theta}_n=\delta_{\overline{x}_n}\otimes\delta_{\overline{y}_n}$, the convexity of the game implies that
$\rho(\overline{\theta}_n) \subset \rho (\widehat{\theta}_n)$ and thus
$\rho^{-1}(E)$ is also approachable in the sense of
$\widetilde{\Gamma}$. The use of displacement approachability
provides however explicit and optimal bounds (see Theorem
\ref{theobhatsetapproch} below). This is the reason we investigate
this special case.

In this framework, we use  proximal gradient normals to define  a
$\widehat{B}$-set:
\begin{definition}\label{Bhatset}
A  closed subset $\widehat{E} \subset \Delta\left(X \times
\Chi\right)$ is a $\widehat{B}$-set if for every $\theta$ not in
$\widehat{E}$ there exist a projection $\underline{\theta} \in
\Pi_{\widehat{E}}(\mu)$, $\overline{p} \in
NP^g_{\widehat{E}}(\underline{\theta})$  and $x=x(\theta) \in
\Delta\left(X\right)$ such that for every $y \in \Chi$, there exists
an optimal plan $\gamma(x,y) \in
\Pi(\underline{\theta},\delta_x\otimes\delta_y)$ and $p(x,y) \in
\mathcal{P}(\gamma(x,y))$ such that:
\[\langle\overline{p},p(x,y)\rangle_{L_2(\underline{\theta})}\leq 0.\]
\end{definition}
This notion of $\widehat{B}$-set extends Blackwell's
one to $\widehat{\Gamma}$ because of the following Theorems
\ref{theobhatsetapproch} and \ref{theoconvexhatapproch}.
\begin{theorem}\label{theobhatsetapproch}
A set $\widehat{E}$ is approachable in $\widehat{\Gamma}$ if and
only if it contains a $\widehat{B}$-set. Given a $\widehat{B}$-set,
the strategy described by $x_{n+1}=x(\widehat{\theta}_n)$ ensures
that
$W_2\left(\widehat{\theta}_n,\widehat{E}\right) \leq K/\sqrt{n}$, for some $K>0$
 \end{theorem}
\textbf{Proof:} Assume that Player~1 plays, at stage $n$,
$x_{n}=x(\widehat{\theta}_{n-1})$ and denote by
$\theta_{n}=\delta_{x_{n}}\otimes\delta_{y_{n}}$ the outcome at
stage $n$. For every $n \in \mathds{N}$, the
displacement average outcome is
$\widehat{\theta}_n=\delta_{\overline{x}_n}\otimes
\delta_{\overline{y}_n}=\delta_{(\overline{x}_n,\overline{y}_n)}$.

If we denote by $\underline{\theta}_n \in \widehat{E}$ the
projection of $\widehat{\theta}_n$ on $\widehat{E}$, then the
optimal plan from $\underline{\theta}_n$ to $\widehat{\theta}_n$ is
$\underline{\theta}_n \otimes \widehat{\theta}_n$. So the proximal
normal  $\overline{p}_n \in NP_{\widehat{E}}(\underline{\theta}_n)$
is defined by
$\overline{p}_n(z)=z-\left(\overline{x}_n,\overline{y}_n\right)$.
Similarly, $\underline{\theta}_n\otimes\left(\delta_{x_{n+1},y_{n+1}}\right)$  is an optimal plan from $\underline{\theta}_n$ to $\delta_{x_{n+1}}\otimes \delta_{y_{n+1}}$, thus if we define $p_{n+1}(z)=z-\left(x_{n+1},y_{n+1}\right)$, the assumption that $\widehat{E}$ is a $\widehat{B}$-set (along
with the choice of $x_{n+1}=x(\widehat{\theta}_n)$) ensures that $\langle
\overline{p}_n,p_{n+1}\rangle_{\underline{\theta}_n} \leq 0$.

As usual, we note that
$W_2^2(\widehat{\theta}_{n+1},\widehat{E})\leq
W_2^2(\widehat{\theta}_{n+1},\underline{\theta}_n)$ which satisfies:
\begin{align*}
W_2^2\left(\widehat{\theta}_{n+1},\underline{\theta}_n\right)&= \int_{(X \times \Chi)^2} \left\|x-z\right\|^2 \mathrm d \widehat{\theta}_{n+1}\otimes \underline{\theta}_n = \int_{X \times \Chi} \left\|\left(\overline{x}_{n+1},\overline{y}_{n+1}\right)-z\right\|^2\mathrm d \underline{\theta}_n(z)\\
&= \int_{X \times \Chi}\left\|\frac{n}{n+1}\left(\overline{x}_{n},\overline{y}_{n}\right)+\frac{1}{n+1}\left(x_{n+1},y_{n+1}\right)-z\right\|^2\mathrm d \underline{\theta}_n(z)\\
&=\left(\frac{n}{n+1}\right)^2\int_{X \times \Chi}\left\|\left(\overline{x}_{n},\overline{y}_{n}\right)-z\right\|^2\mathrm d \underline{\theta}_n(z)\\&+\left(\frac{1}{n+1}\right)^2\int_{X \times \Chi}\left\|\left(x_{n+1},y_{n+1}\right)-z\right\|^2\mathrm d \underline{\theta}_n(z)\\
&+2\frac{n}{(n+1)^2}\int_{X \times \Chi}\left\langle\left(\overline{x}_{n},\overline{y}_{n}\right)-z,\left(x_{n+1},y_{n+1}-z\right)\right\rangle\mathrm d \underline{\theta}_n(z).
\end{align*}
Therefore,
\begin{align*}
W_2^2\left(\widehat{\theta}_{n+1},\underline{\theta}_n\right)&=\left(\frac{n}{n+1}\right)^2W_2^2\left(\widehat{\theta}_n,\underline{\theta}_n\right)+\left(\frac{1}{n+1}\right)^2W_2^2\left(\theta_{n+1},\underline{\theta}_n\right)\\&+2\frac{n}{(n+1)^2}\langle
\overline{p}_n,p_{n+1}\rangle_{\underline{\theta}_n} \leq
\left(\frac{n}{n+1}\right)^2W_2^2\left(\widehat{\theta}_n,\widehat{E}\right)+\left(\frac{K}{n+1}\right)^2.
\end{align*}
We conclude by induction over $n \in \mathds{N}$.

\bigskip

We sketch the proof of the necessary part. Conclusions of Lemma
\ref{lemmeSpinat} (delayed to Appendix) hold in $\widehat{\Gamma}$ and the proof of the
first two points are identical.  Hence it remains to prove the third
point, i.e.\ that a set which is not a $\widehat{B}$-set has a
secondary point (see Definition \ref{defsecondary} also in Appendix). Let $\overline{\theta}$ be not in $\widehat{E}$,
$\underline{\theta}$ one of its projection on $\widehat{E}$, and
$\overline{p} \in \NC_{\widehat{E}}^g(\underline{\theta})$ the
associated proximal normals such that:
\[\forall x \in X, \exists y \in \Chi, \langle \underline{p},p(x,y) \rangle_{\underline{\theta}} = \int_{X \times \Chi} \langle \overline{p}(z),z-(x,y)\rangle \mathrm d \underline{\theta} >0.\]
Sion's theorem implies the existence of $\delta>0$ and  $y \in \Chi$ such that
for every $x \in X$,  $\langle \underline{p},p(x,y)
\rangle_{\underline{\theta}} \geq \delta$. If we denote by
$\theta_{\lambda}=\left(\Id,\sigma_{\lambda}\right)\sharp
\underline{\theta}_n\otimes \theta_{n+1}$ then using the same
argument as in the proof of Lemma \ref{convex-Cond}, we show that
\[W_2(\overline{\theta},\theta_{\lambda}) \leq W_2(\overline{\theta},\underline{\theta})+\frac{K\lambda^2-2 \lambda \delta}{2W_2(\overline{\theta},\underline{\theta})}\leq W_2(\overline{\theta},\underline{\theta})-\frac{\lambda \delta}{2W_2(\overline{\theta},\underline{\theta})},\]
for $\lambda$ small enough. Hence, $\underline{\theta}$ is a
secondary point.  $\hfill \Box$

\bigskip

The following Theorem is  the characterization of displacement
convex approachable sets.
\begin{theorem}\label{theoconvexhatapproch}
A displacement convex set $\widehat{C}$ is approachable by Player~1
in $\widehat{\Gamma}$ if and only if
\[ \forall\, y \in \Chi, \exists\, x \in X,\  \theta(x,y)=\delta_x\otimes\delta_y \in
\widehat{C}.\]
\end{theorem}
The proof is based on the following lemma that extends  to Wasserstein space the usual characterization of the projection on a convex set in an Euclidian space .
\begin{lemma} \label{convex-Cond}
Let $X$ be a compact subset of  $\mathds{R}^N$ and $A$ be a
displacement   convex subset of $\Delta(X)$. Fix $\underline{\theta}
\in A$. Then for all $ \overline{p} \in \NC^g_A (\underline{\mu}) $
and all $ \theta_1 \in A$ we have
 \begin{equation}  \label{pdtscal}  \forall p \in \mathcal{P} (\underline{\theta} ,  \theta_1 ), \;\;
\int _{\mathds{R} ^N } \langle\overline{p} (x), p(x)\rangle \mathrm
d \underline{\theta} (x) :=  \langle \overline{p} ,
p\rangle_{\underline{\theta}} \leq 0.
 \end{equation}
 \end{lemma}
\textbf{Proof:} Let us consider  $\underline{\theta},\theta_0 \in A$
and $ \overline{p} \in NP_{A}^g(\underline{\theta}) $. We denote by $ \theta
\notin A$ the measure outside $A$  and $ \gamma  \in \Pi (\underline{\theta} , \theta) $ the optimal plan given by the definition of proximal gradient normals. Define
$\gamma ' = T \sharp \gamma $ where $T : (x,y) \mapsto (y,x)$ so that  $\gamma' $ is obviously an optimal plan from $\theta $ to $
\underline{\theta}$.

Let  $ \widetilde{\gamma} \in \Pi ( \underline{\theta}, \theta_0) $
be an optimal plan from $ \underline{\theta} $ to $ \theta_0 $ and
for any $ \lambda \in [0,1] $ we  define $ \theta_{\lambda} :=
\sigma_{\lambda} \sharp \widetilde{\gamma}$  and
$\widetilde{\gamma}_{\lambda}=(\Id,\sigma_{\lambda})\sharp
\widetilde{\gamma}$ which belongs respectively to the displacement
convex set $A$ and to $\Pi(\underline{\theta},\theta_{\lambda})$.

By the disintegration of measure theorem  for any $y \in
\mathds{R}^N $ there exists a probability  measure $
\widetilde{\gamma}_{\lambda,y} $ on $\mathds{R}^{N}$ such that $
\widetilde{\gamma}_{\lambda}   = \int _{\mathds{R} ^N} ( \delta_y
\otimes \widetilde{\gamma}_{\lambda,y}  ) \underline{\theta}
(\mathrm dy)$ which means that for any continuous bounded function
$u (y,z) :\;  \mathds{R} ^{2N} \mapsto \mathds{R} $
\[\int _ {\mathds{R} ^{2N} } u(y,z) \widetilde{\gamma}_{\lambda} (\mathrm dy,\mathrm dz) = \int _{\mathds{R}^N }\left[\int  _{\mathds{R}^N}  u(y,z)  \widetilde{\gamma}_{\lambda,y} ( \mathrm dz) \right] \underline{\theta} (\mathrm dy)  \]

 We define $\widehat{\gamma} \in \Pi(\theta,\theta_{\lambda})$ by:
\[\forall \phi \in C_b, \int_{\mathds{R}^{2N}}\phi \mathrm d\widehat{\gamma}=  \int_{\mathds{R}^{3N}}\phi(x,z)\widetilde{\gamma}_{\lambda,y}(\mathrm dz)\gamma'(\mathrm dx,\mathrm dy).\]
Since $\theta_{\lambda} \in A$ and $\widehat{\gamma} \in
\Pi(\theta,\theta_{\lambda})$, we obtain:
\begin{align*}W_2^2 (\theta , \underline{\theta}) &\leq W_2^2(\theta,\theta_{\lambda})
\leq \int_{\mathds{R}^{2N}}\|x-z\|^2\mathrm d \widehat{\gamma}\\
&= \int_ {\mathds{R}^{3N} } \|x-z\|^2\widetilde{\gamma}_{\lambda,y}(\mathrm dz)\gamma'(\mathrm dx,\mathrm dy)\\
& =  \int _ {\mathds{R}^{3N} }  \|x-y \|^2
\widetilde{\gamma}_{\lambda,y} (\mathrm dz)  \gamma '(\mathrm
dx,\mathrm dy) + 2\int _ {\mathds{R}^{3N} }  \langle x-y,y-z\rangle
\widetilde{\gamma}_{\lambda,y} (\mathrm dz)  \gamma '(\mathrm
dx,\mathrm dy)
\end{align*} \[+\int _ {\mathds{R}^{3N} } \|y-z \|^2  \widetilde{\gamma}_{\lambda,y} (\mathrm dz)  \gamma '(\mathrm dx,\mathrm dy) = a + b+ c \]  where $a$, $b$ and $c$ denote
respectively the three integral terms in the above equality. It remains to estimate the three terms $a$, $b$ and $c$.
\[ a =  \int _ {\mathds{R}^{2N} } \|x-y \|^2   \gamma '(\mathrm dx,\mathrm dy) = W_2 ^2 ( \theta , \underline{\theta} ).\]
\begin{align*} b &=  2\int _ {\mathds{R}^{2N} } \left\langle x-y, \int _ {\mathds{R}^N } (y-z ) \widetilde{\gamma}_{\lambda,y} (\mathrm dz) \right\rangle \gamma ' (\mathrm dx,\mathrm dy )\\
&=  2\int _ {\mathds{R}^{2N} } \left\langle y-x, \int _ {\mathds{R}^N } (x -z ) \widetilde{\gamma}_{\lambda,x} (\mathrm dz) \right\rangle  \gamma   (\mathrm dx,\mathrm dy)  \quad \mbox{  (by definition de $ \gamma ' $) } \\
 &=  - 2\int _ {\mathds{R}^{N} } \left\langle\overline{p} (x) , \int _ {\mathds{R}^N } (x -z ) \widetilde{\gamma}_{\lambda,x} (\mathrm dz) \right\rangle  \underline{\theta}  (\mathrm dx) \qquad \mbox{ (from the definition of $\overline{p}$) }  \\
&= - 2\int _ {\mathds{R}^{2N} } \left\langle\overline{p} (x) , x -z \right\rangle  \widetilde{\gamma}_{\lambda,x} (\mathrm dz)\underline{\theta}  (\mathrm dx) \\
& = - 2\int _ {\mathds{R}^{2N} } \left\langle\overline{p} (x) , x -z
\right\rangle  \widetilde{\gamma}_{\lambda}  (\mathrm dx,\mathrm dz)
\qquad \mbox{  (by the desintegration formula) }\\
&=-2\int _ {\mathds{R}^{2N} } \left\langle\overline{p} (x) , x -\left[(1-\lambda)x+\lambda z\right]  \right\rangle  \widetilde{\gamma}  (\mathrm dx,\mathrm dz) \qquad \mbox{  (by definition of } \widetilde{\gamma}_{\lambda}) \\
&=-2\lambda\int _ {\mathds{R}^{2N} } \left\langle\overline{p} (x) ,
x -z  \right\rangle \widetilde{\gamma}  (\mathrm dx,\mathrm dz)=  -2
\lambda \int _ {\mathds{R}^{2N} } \left\langle\overline{p} (x) ,
p(x)  \right\rangle \underline{\theta}  (\mathrm dx). \end{align*}
And this holds for any $p \in \mathcal{P} (\underline{\theta}, \theta_0)$.

The disintegration of measure formula together with the definition
of $ \widetilde{\gamma}$ yield
\[c  =   \int _ {\mathds{R}^{2N} } \big\| y - \left[(1-\lambda)y+\lambda z\right] \big\|^2  \mathrm d\widetilde{\gamma} (y,z)  = \lambda ^2 \int _ {\mathds{R}^{2N} } \big\| y - z \big\|^2  \mathrm d\widetilde{\gamma} (y,z)\]
hence $c=\lambda^2W_2^2(\underline{\theta},\theta_0)$.

Summarizing our estimates, we have obtained
\[ W_2 ^2 ( \theta, \underline{\theta}) \leq  W_2 ^2 ( \theta, \underline{\theta}) - \lambda  \int _ {\mathds{R}^{2N} }2 <\overline{p} (x) , p(x)  >  \underline{\theta}  (\mathrm dx) +\lambda ^2W_2^2(\underline{\theta},\theta_0).\]

Thus for any $\lambda \in (0, 1)$,
\[ 0 \leq  \lambda ^2 W( \underline{\theta},  \theta_1) - 2\lambda  \int _ {\mathds{R}^{2N} }<\overline{p} (x) , p(x)  >  \underline{\theta}  (\mathrm dx).\]
Dividing firstly  by $\lambda  >0$  and letting secondly   $ \lambda
$ tend to $0 ^+ $, this gives the wished conclusion.  $\hfill \Box$

\bigskip

\textbf{Proof of Theorem \ref{theoconvexhatapproch}.} Let $\theta$
be any measure not in $\widetilde{E}$ and denote by
$\underline{\theta} \in \Pi_{\widehat{C}}(\theta)$ any of its
projection and $\overline{p} \in
NP_{\widehat{C}}(\underline{\theta})$, associated to some
$\overline{\gamma} \in \Pi(\underline{\theta},\theta)$, any proximal
normal. For every $x \in X$ and $y \in \Chi$  the only optimal plan
from $\underline{\theta}$ to $\delta_x\otimes \delta_y$ is
$\widehat{\gamma}=\underline{\theta}\otimes(\delta_x\otimes\delta_y)$.

 The function $h:X \times
\Chi$ defined by
\[h(x,y)=\int_{\left(X \times \Chi\right)^2}<\underline{p}(u),u-v>\mathrm d\widehat{\gamma}(x,y)\]
is affine in both of its variable since:
\[h(x,y)=\int_{X \times
\Chi}\left\langle\overline{p}(u),u\right\rangle\mathrm
d\underline{\theta}(u)-\left\langle\int_{X \times
\Chi}\overline{p}(u)d\underline{\theta}(u),\left(x,y\right)\right\rangle=\overline{z}-\langle
z,\left(x,y\right)\rangle ,\]
\[ \mathrm{\ where \ } \overline{z}= \int_{X \times
\Chi}\langle\overline{p}(u),u\rangle\mathrm d\underline{\theta}(u)
\mathrm{\ and \ } z=\int_{X \times
\Chi}\overline{p}(u)d\underline{\theta}(u).\]

Since for every $y \in \Chi$, there exists $x \in X$ such that
$\delta_x\otimes \delta_y \in C$, Proposition \ref{convex-Cond}
implies that for every $y \in \Chi$ there exists $x \in X$ such that
$h(x,y)\leq0$. $X$ and $\Chi$ are compact sets, therefore Sion's
theorem implies that there exists $x \in X$ such that $h(x,y)$ for
every $y \in \Chi$. Hence $\widehat{C}$ is a $\widehat{B}$-set and
is approachable by Player~1.

Reciprocally, assume that there exists $y\in \Chi$ such that
$\delta_x \otimes \delta_y \not \in \widehat{C}$ for every $x \in
X$. Since $X$ is compact, there exists $\eta>0$ such that  $\inf_{x
\in X}W_2\left(\delta_x\otimes\delta_y, \widehat{C}\right)\geq
\eta$. The strategy of Player~2 that consists of playing at each
stage $\delta_y$ ensures that
$\widetilde{\theta_n}=\delta_{\overline{x}_n}\otimes\delta_y$ is
always at, at least, $\delta>0$ from $\widehat{C}$. Therefore it is
not approachable by Player~1. $\hfill \Box$
\section*{Concluding remarks}
Recall that  action spaces in  $\widetilde{\Gamma}$ (resp.\
$\widehat{\Gamma}$) are  $\Delta(X)$ and $\Delta(\Chi)$ (resp.\ $X$
and $\Chi$). Assume now that in  $\widetilde{\Gamma}$ players are
restricted to $X$ and $\Chi$; then a  $\widetilde{B}$-set should
satisfy:
\[\forall \underline{\theta} \in \widetilde{E},\, \forall \phi \in \NC_{\widetilde{E}}^p(\underline{\theta}),\, \exists x \in X,\, \forall y \in \Chi,\ \int_{X \times \Chi} \phi  \ \mathrm d\left(\underline{\theta}- \delta_x\otimes\delta_{y}\right) \leq 0.\]
The proof of the sufficient part of Theorem \ref{tildecond} does not
change when we add this assumption, thus a $\widetilde{B}$-set is
still approachable. However, both the proof of  the necessary part
of Theorem \ref{tildecond} and the proof of Theorem
\ref{tildeconvexe} are no longer valid (due to the lack of
linearity).

\medskip

Similarly, assume that in $\widehat{\Gamma}$ players can choose
action in $\Delta(X)$ and $\Delta(\Chi)$ and,  at stage $n \in
\mathds{N}$, the outcome is $\theta_n=\mathbf{x}_n \otimes
\boldsymbol{\xi}_n$. Strictly speaking,  given such outcomes that
might not be absolutely continuous with respect to
$\boldsymbol{\lambda}$, the sequence of interpolation
$\widehat{\theta}_n$ may not be unique. However, we can assume that
the game begins at stage 2 and that $\theta_1 =
\boldsymbol{\lambda}/\boldsymbol{\lambda}(X\times \Chi)$; then, see
e.g.\ Villani~\cite{Villani}, Proposition 5.9,  $\widehat{\theta}_2
\ll \boldsymbol{\lambda}$ and is  unique.  By induction, the
sequence of $\widehat{\theta}_n$ is unique. Once again, using the
same proof, we can show that a $\widehat{B}$-set is displacement
approachable, but we cannot extends the necessary part nor the
characterization of  displacement approachable convex sets.

\appendix

\section{Proof of Proposition \ref{tildenormalequiv}} Let us first state and prove the following (implicitly stated) lemma:

\begin{lemma}\label{lemmalienW2}
If the two functions $\rho$ and $s$ are linear both in $x \in \Delta(I)$ and $\mu \in \mathcal{S}$, then the multivalued mapping $(x,\mu) \mapsto
P(x,\mu)=\left\{\rho(x,y);\ y \in \mathbf{s}^{-1}(\mu)\right\}$ is a
$L$-Lipschitzian convex hull.
\end{lemma}
\textbf{Proof:} Since  the graph  of $\mathbf{s}^{-1}$ is  a
polytope  of $\mathds{R}^{SI}\times \mathds{R}^{J}$, there exists a
finite family of   (so called) \textit{extreme points} functions
$\{y_\kappa(\cdot);\ \kappa \in \mathcal{K}\}$ from $\mathcal{S}$
into $\Delta(J)$ that are all piecewise linear and  continuous (thus
Lipschitzian)  such that
$\mathbf{s}^{-1}(\mu)=\co\left\{y_{\xi}(\mu);\ \kappa \in
\mathcal{K}\right\}$, for every $\mu \in \mathcal{S}$. Since
$\rho(x,\cdot)$ is linear on $\Delta(J)$:
\[P(x,\mu)=\left\{\rho(x,y);\ y \in \mathbf{s}^{-1}(\mu)\right\}= \left\{\rho(x,y);\ y \in \co\left\{y_{\kappa}(\mu), \kappa \in \mathcal{K}\right\}\right\}
= \co \bigg\{ \rho(x,y_{\kappa}(\mu));\ \kappa \in \mathcal{K}
\bigg\}.\] Therefore $P$ is indeed a $L$-Lipschitzian
convex hull. $\hfill \Box$
\medskip

We now turn to the actual proof of Proposition \ref{tildenormalequiv}:

\textbf{Proof of Proposition \ref{tildenormalequiv}}

The third point is obvious, so we only need to prove
that if $\widetilde{E} \subset \Delta\left(X\times \Chi\right)$ is
approachable in $\widetilde{\Gamma}$ then
$\rho(\widetilde{E})\subset \mathds{R}^k$ is approachable in
$\Gamma$ (see \textbf{part 1}) and that if $E \subset
\mathds{R}^k$ is approachable in $\Gamma$
 then $\rho^{-1}(E) \subset \Delta\left(X\times \Chi\right)$ is also approachable (see \textbf{part 2}). The remaining easily follows from the fact that $\rho\big(\rho^{-1}(E)\big) \subset E$.

\medskip

\textbf{part 1:} The proof consists in two steps. First, we link the
Wasserstein distance between two probability measures
$\overline{\theta}, \underline{\theta} \in \Delta(X \times \Chi)$
and the distance between the two sets
$\rho(\overline{\theta})\subset \mathds{R}^k$ and
$\rho(\underline{\theta}) \subset \mathds{R}^k$. We will prove this step
with the use of the 1-Wasserstein distance defined below. In the
second step, we transform a strategy in $\widetilde{\Gamma}$ into a
strategy in $\Gamma^p$.

\bigskip

\underline{Step 1:} The 1-Wasserstein distance between $\mu$ and
$\nu$ in $\Delta(X)$ is defined by:
\[W_1(\overline{\theta},\underline{\theta})=\inf_{\gamma \in \Pi(\overline{\theta},\underline{\theta})} \int_{X}\|x-y\|\mathrm d\gamma(x,y)=\sup_{\phi \in \LIP_1(X,\mathds{R})} \int_X \phi\ \mathrm d\overline{\theta} - \mathrm d\underline{\theta} = \inf_{U\sim \overline{\theta}, V \sim \underline{\theta}} \mathds{E}[\|U-V\|],\]
where $\LIP_1(X,\mathds{R})$ is the set of 1-Lipschitzian functions
from $X$ to $\mathds{R}$. Jensen's inequality and the  probabilistic
interpretation imply that $W_1(\overline{\theta},\underline{\theta})
\leq W_2(\overline{\theta},\underline{\theta})$.

Let $P$ is a $L$-Lipschitzian convex hull, then since  $P(x,\xi)=\co
\big\{p_{\kappa}(x,\xi); \kappa \in \mathcal{K}\big\}$ where every
$p_{\kappa}$ is $L$-Lipschitzian, for every $\overline{\theta} \in
\Delta(X \times \Chi)$, by convexity of the integral (see e.g.\ Klein
and Thompson~\cite{Kleinthompson}, Theorem 18.1.19):
 \[\int_{X \times \Chi} P(x,\xi) \mathrm d\overline{\theta}= \int_{X \times \Chi} \co\left\{p_{\kappa}(x, \xi);\ \kappa \in \mathcal{K}\right\}\mathrm d\overline{\theta}=\co \left\{\int_{X \times \Chi} p_{\kappa}(x, \xi) \mathrm d\overline{\theta};\ \kappa \in \mathcal{K}\right\}.\]The mapping $p_{\kappa}$ is $L$-Lipschitzian, so $d\left(\int_{X \times \Chi}
p_{\kappa}(x, \xi) \mathrm d\overline{\theta},
\rho(\underline{\theta})\right)\leq \sqrt{k}L\varepsilon$ and since the set
$\rho(\underline{\theta})$ is  convex,
$d\left(\rho(\overline{\theta}),\rho(\underline{\theta})\right) \leq
\sqrt{k}L\varepsilon$.  Therefore,
\[ W_2\left(\overline{\theta},\widetilde{E}\right)\leq \varepsilon \ \Longrightarrow \ \sup_{z
\in \rho(\overline{\theta})} d(z,\rho(\widetilde{E})) \leq
\sqrt{k}L\varepsilon\, .\]

 \medskip

\underline{Step 2:} This step transforms a strategy in $\widetilde{\Gamma}$ into a strategy in $\Gamma$ and is quite standard  in games with partial monitoring (see e.g.\ Lugosi, Mannor and Stoltz \cite{LugosiMannorStoltz}); its proof, which relies deeply Hoeffding-Azuma's~\cite{Hoeffding,Azuma} inequality, is therefore only sketched.

Let $\widetilde{\sigma}$ be a strategy of Player~1 that approaches
(up to $\varepsilon>0$) a set $\widetilde{E} \subset
\Delta\left(X\times \Chi\right)$. This strategy cannot be directly
played in $\Gamma$ in order to approach $\rho(\widetilde{E})$ for
two reasons:
\begin{itemize}
\item[1)]{in $\Gamma$, Player~1 chooses an action $i \in I$ and not some $\mathbf{x} \in \Delta(\Delta(I))$;}
\item[2)]{at stage $n$ in $\Gamma$,  the flag $\mu_n=\mathbf{s}(i_n,j_n)$ is not observed, but  only  a signal $s_n$ whose law is $s(i_n,j_n)$.}
\end{itemize}
The usual trick is to divide $\mathds{N}$ into  blocks of length $N \in \mathds{N}$ -- where $N$ is big enough. The $n$-th block in $\Gamma$ will
correspond to the $n$-th stage in $\widetilde{\Gamma}$ and $\sigma$ is defined inductively.  Assume that  $\widetilde{\sigma}$  dictates to play  $\mathbf{x}_n \in \Delta(X)$  at the $n$-th stage of $\widetilde{\Gamma}$.  In $\Gamma$ and independently at every stage $t$ of the $n$-th block, with probability $\eta$  the action $i_t$ is chosen uniformly over $I$, and with probability $1-\eta$ accordingly to $\mathds{E}_{\mathbf{x}_n} \in X$.

The stages where $i_t$ was chosen uniformly allow to build an unbiased estimator that will be arbitrarily close (if $N$ is big enough) to  $\overline{\mu}_n$, the average flag during the $n$-th block. Since the choice of actions are independent, it is easy to show that the average payoff on the $n$-th block is arbitrarily close to the $\eta$-neighborhood of $P(\mathds{E}_{\mathbf{x}_n},\overline{\mu}_n)=\rho({E}_{\mathbf{x}_n},\delta_{\overline{\mu}_n})$. So it is enough to act as if  Player 2's action in $\widetilde{\Gamma}$ was $\delta_{\overline{\mu}_n}$.

In order to obtain the almost surely convergence, we  can use a classical {\em doubling trick} argument (see e.g. Sorin \cite{SorinSupergames}), which  consists in a concatenation of these  strategies with increasing $N$ and decreasing $\eta$.

\bigskip

\textbf{Part 2:} Assume that $E \subset \mathds{R}^k$ is
approachable in $\Gamma$ by Player~1. Consider the game  where
Player~1 observes in addition $\mu_n=\mathbf{s}(j_n)$ and his payoff
is  $\mathds{E}_{x_n}[\rho(i_n,j_n)]$ where $x_n$ is the law of
$i_n$. This new game is {\em easier} for Player~1 because he has
more information and actions, hence he can still approach $E \subset
\mathds{R}^k$. Since $P(x,\cdot)$ is convex,
allowing  Player~2 to play any action in $\Delta(\Chi)$ does not
make the game harder for Player~1. Thus  we can assume that at stage
$n \in \mathds{N}$,  Player~1 observes $\boldsymbol{\xi}_n \in
\Delta(\Chi)=\Delta(\mathcal{S})$, that  he plays deterministically
$\mathbf{x}_n \in \Delta(X)=\Delta(\Delta(I))$ and that his payoff
belongs to $\rho(\mathbf{x}_n\otimes\boldsymbol{\xi}_n)$. We call this new
game by $\Gamma^d$.

The fact that $E$ is approachable in $\Gamma^d$ implies that for
every $\varepsilon$ there exists a strategy $\sigma_{\varepsilon}$
in $\Gamma^d$ and $N_{\varepsilon} \in \mathds{N}$ such that for
every $n \geq N_{\varepsilon}$ and strategy $\tau$ of Player~2:
 \begin{equation}\label{equapart21} d\left(\frac{\sum_{m=1}^n\rho\left(\mathbf{x}_m\otimes\boldsymbol{\xi}_m\right)}{n},E\right):=\sup \left\{d(z,E);\ z \in \frac{\sum_{m=1}^n\rho\left(\mathbf{x}_m\otimes\boldsymbol{\xi}_m\right)}{n}\right\} \leq \varepsilon.\end{equation}
If we denote as before $\theta_n=\mathbf{x}_n\otimes
\boldsymbol{\xi}_n \in \Delta(X \times \Chi)$ then equation
(\ref{equapart21}) becomes:
 \[\forall \varepsilon>0,\, \exists N_{\varepsilon} \in \mathds{N},\, \forall n \geq N_{\varepsilon},\ \overline{\theta}_n \in \rho^{-1}(E^{\varepsilon}).\]

 Let us  define similarly $\rho^{-1}(E)^{\delta}=\left\{\theta \in \Delta\left(X \times \Chi\right);\ W_2\left(\theta,\widetilde{E}\right)\leq \delta\right\}$.  Since the sequence of compact sets $\rho^{-1}\left(E^{\varepsilon}\right)$  converges (as $\varepsilon$ converges to zero) to $\rho^{-1}(E)$, for every $\delta>0$, there exists $\underline{\varepsilon}$ such that for every $0<\varepsilon \leq \underline{\varepsilon}$,  $\rho^{-1}\left(E^{\varepsilon}\right)\subset\rho^{-1}(E)^{\delta}$. Therefore, for every $\delta>0$, there exists $N \in \mathds{N}$ such that for every $n \geq N$ and every strategy $\tau$ of Player~2, $\overline{\theta}_n$ belongs to $\rho^{-1}(E)^{\delta}$. Thus $\rho^{-1}(E)$ is approachable by Player~1.
  $\hfill \Box$

\section{Proof of the necessary part of Theorem \ref{tildecond}}

The necessary part of  Theorem \ref{tildecond} is an immediate consequence of Lemma \ref{lemmeSpinat}, greatly inspired from Spinat \cite{Spinat}; it requires the  following definition.

\begin{definition}\label{defsecondary}
A point $\theta \in \Delta(X \times \Chi)$ is $\delta$-secondary for
$\widetilde{E}$ if there exists a corresponding couple: a point
$\boldsymbol{\xi} \in \Delta(\Chi)$ and a continuous function
$\lambda: \Delta(X) \to (0,1]$ such that $\min_{\mathbf{x} \in
\Delta(X)}
W_2\left(\lambda(\mathbf{x})\theta+(1-\lambda(\mathbf{x}))\mathbf{x}\otimes\boldsymbol{\xi},\widetilde{E}\right)\geq
\delta$. A point $\theta$ is secondary to $\widetilde{E}$ if there
exists $\delta>0$ such that $x$ is $\delta$-secondary to
$\widetilde{E}$.

We denote by $\mathcal{PP}(\widetilde{E})\subset \widetilde{E}$ the
subset of primary point to $\widetilde{E}$ (i.e.\ points of
$\widetilde{E}$ that are not secondary).
\end{definition}

\begin{lemma}[Spinat~\cite{Spinat}]\label{lemmeSpinat}\
\begin{enumerate}
\item[i)]{Any approachable compact set contains a minimal approachable set;}
\item[ii)]{A minimal approachable set is a fixed point of $\mathcal{PP}$;}
\item[iii)]{A fixed point of $\mathcal{PP}$ is a $\widetilde{B}$-set.}
\end{enumerate}
\end{lemma}
\textbf{Proof:} \textit{i)}  Let
$\mathcal{B}=\left\{\left.\widetilde{B} \subset \widetilde{E}
\right| \widetilde{B} \mathrm{ \ is \ an \ approachable \ compact \
set}\right\}$ be a nonempty family ordered by inclusion. Every fully
ordered subset of  $\mathcal{B}$ has a minorant $\underline{\widetilde{B}}$ (the
intersection of every elements of the subset) that belongs to
$\mathcal{B}$ since it is an approachable compact subset of
$\widetilde{E}$. Thus Zorn's lemma yields that $\mathcal{B}$
contains at least one minimal element.

\textit{ii)} We claim that if $\widetilde{E}$ is approachable then
so is $\mathcal{PP}(\widetilde{E})$, hence a minimal approachable set
is necessarily a fixed point of $\mathcal{PP}$. Indeed, if $\theta$
is $\delta$-secondary  there exists an open neighborhood $V$ of
$\theta$ such that  every point of $V$ is $\delta/2$-secondary,
because of the continuity of $W_2$. Hence
$\mathcal{PP}(\widetilde{E})$ is a compact subset of $\widetilde{E}$.

Let $\theta_0$ be a $\delta$-secondary point of an approachable set
$\widetilde{E}$ and $\boldsymbol{\xi}, \lambda$ the associated
couple given in Definition \ref{defsecondary}. Let $\varepsilon <
\delta/4$ and consider $\sigma$ a strategy of Player~1 that ensures
that $\overline{\theta}_n$ is, after some stage $N \in \mathds{N}$,
closer than $\varepsilon$ to $\widetilde{E}$. We will show that
$\overline{\theta}_n$ must be  close  to $\theta_0$  only a finite
number of times; so  Player~1 can approach
$\widetilde{E}\backslash\{\theta\}$. Indeed, assume that there
exists a  stage $m \in \mathds{N}$ such that
$W_2(\overline{\theta}_m, \theta_0) \leq \delta/4$ and consider the
strategy of Player~2 that consists in playing repeatedly
$\boldsymbol{\xi}$ from this stage on. It is clear that (if $m$ is
big enough) after some stage $\overline{\theta}$ will be
$\delta/2$-closed to
$\lambda\left(\overline{\mathbf{x}}_{n,m}\right)\theta_0+\left(1-\lambda\left(\overline{\mathbf{x}}_{n,m}\right)\right)\overline{\mathbf{x}}_{n,m}\otimes
\boldsymbol{\xi}$ where $\overline{\mathbf{x}}_{n,m}$ is the average
action played by Player~1 between stage $m$ and $m+n$. Therefore
$W_2\left(\overline{\theta}_n,\widetilde{E}\right) \geq
\delta/2>\varepsilon$  and since  $W_2(\overline{\theta}_n,
\theta_0)$ can be bigger  than $\delta/4$ only a finite number of
times,  the strategy of Player~1  approaches
$\widetilde{E}\backslash\{\theta\}$. This is true for any
secondary point, so Player~1 can approach
$\mathcal{PP}(\widetilde{E})$.

\textit{iii)} Assume that $\widetilde{E}$ is not a
$\widetilde{B}$-set: there exists $\overline{\theta} \not \in
\widetilde{E}$ such that for any projection $\underline{\theta} \in
\Pi_{\widetilde{E}}(\overline{\theta})$, any $\phi \in
\Phi(\underline{\theta},\overline{\theta})$ and any $\mathbf{x} \in
\Delta(X)$, there exists $\boldsymbol{\xi} \in \Delta(\Chi)$ such
that  $\int_{X \times \Chi}\phi \mathrm d(\underline{\theta}
-\mathbf{x}\otimes\boldsymbol{\xi}) >0$. This last expression is
linear both in $\mathbf{x}$ and $\boldsymbol{\xi}$, so  Von
Neumann's minmax theorem imply that there exists
$\boldsymbol{\xi}(=\boldsymbol{\xi}(\underline{\theta},\phi))$ and
$\delta$ such that $\int_{X \times \Chi}\phi \mathrm
d(\underline{\theta}- \mathbf{x}\otimes\boldsymbol{\xi}) \geq
\delta>0$, for every $\mathbf{x} \in \Delta(X)$.

We can assume that $\overline{\theta} \in \Delta_{0}(X\times\Chi)$. Otherwise, let
$(\overline{\theta}_n \in \Delta_{\frac{1}{n}}(X\times\Chi))_{n \in \mathds{N}}$ be   a sequence of
measures that converges to $\overline{\theta}$ ,
$(\underline{\theta}_n)_{n \in \mathds{N}}$ a  sequence of
projection of $\overline{\theta}_n$ onto $\widetilde{E}$ and
$\phi^n_0\in \Phi(\overline{\theta}_n,\underline{\theta}_n)$. Up to
two extractions, we can assume that $\underline{\theta}_n$ converges
to $\underline{\theta}_0$ a projection of $\overline{\theta}$ and
$\phi_0^n$ converges to $\phi_0 \in
\Phi(\overline{\theta},\underline{\theta}_0)$. Therefore, for $n$
big enough and  for   every $\mathbf{x} \in \Delta(X)$,
\[0<\frac{\delta}{2}\leq \int_{X \times \Chi}\phi_0^n \mathrm d\left(\underline{\theta}_n- \mathbf{x}\otimes\boldsymbol{\xi}(\underline{\theta}_0,\phi_0)\right)\]
since the right member converges to $\int_{X \times
\Chi}\phi_0\mathrm d\left(\underline{\theta}_0-\mathbf{x}\otimes
\boldsymbol{\xi}(\underline{\theta}_0,\phi_0)\right) \geq \delta$.

For every $\lambda \in [0,1]$ and $\mathbf{x} \in \Delta(X)$, we
denote by $\phi_{\lambda,\mathbf{x}}$ the unique (we assumed that $\overline{\theta} \ll_0$)
Kantorovitch potential such that:
\begin{align*}W^2_2\left((1-\lambda)\underline{\theta}+\lambda \mathbf{x}\otimes \boldsymbol{\xi},\overline{\theta}\right)&=\int_{X \times \Chi} \phi_{\lambda,\mathbf{x}} \mathrm d\left((1-\lambda)\underline{\theta}+\lambda \mathbf{x}\otimes \boldsymbol{\xi}\right)+\int_{X \times \Chi} \phi_{\lambda,\mathbf{x}}^*\mathrm d\overline{\theta}\\
&=\int_{X \times \Chi} \phi_{\lambda,\mathbf{x}} \mathrm
d\underline{\theta}+\int_{X \times
\Chi}\phi_{\lambda,\mathbf{x}}^*\mathrm
d\overline{\theta}-\lambda\int_{X \times \Chi}
\phi_{\lambda,\mathbf{x}} \mathrm
d(\underline{\theta}-\mathbf{x}\otimes
\boldsymbol{\xi}).\end{align*} Since $(\lambda,\mathbf{x}) \mapsto
\phi_{\lambda,\mathbf{x}}$ is continuous,
$\phi_{\lambda,\mathbf{x}}$ converges to $\phi_0$, for every
$\mathbf{x} \in \Delta(X)$ which is compact. Hence there exists
$\underline{\lambda} \in (0,1]$ such that:
 \[\left|\int_{X \times \Chi}\left(\phi_{\lambda,\mathbf{x}}-\phi_0\right)\mathrm d(\underline{\theta}-\mathbf{x}\otimes\boldsymbol{\xi})\right|\leq \delta/4, \quad \forall \lambda \leq \underline{\lambda}, \forall \mathbf{x} \in \Delta(X).\]
Therefore, one has $\Big(W_
2\left(\overline{\theta},(1-\underline{\lambda})\underline{\theta}+\underline{\lambda}\mathbf{x}\otimes
\boldsymbol{\xi}\right)\Big)^2\leq
\Big(W_2(\overline{\theta},\underline{\theta})\Big)^2-\underline{\lambda}\frac{\delta}{4}$
so\[W_
2\left(\overline{\theta},(1-\underline{\lambda})\underline{\theta}+\underline{\lambda}\mathbf{x}\otimes
\boldsymbol{\xi}\right)\leq
W_2(\overline{\theta},\underline{\theta})-\frac{\underline{\lambda}\delta}{8W_2(\overline{\theta},\underline{\theta})}:=W_2(\overline{\theta},\underline{\theta})-\eta\]
which implies that
$W_2\left((1-\underline{\lambda})\underline{\theta}+\underline{\lambda}
\mathbf{x}\otimes \boldsymbol{\xi},\widetilde{E}\right) \geq \eta$
and $\underline{\theta}$ is $\eta$-secondary to $\widetilde{E}$.

Consequently, a fixed point of $\mathcal{PP}$, i.e.\ a set without
any secondary point, is necessary a $\widetilde{B}$-set. $\hfill \Box$

\medskip

{\bf Acknowledgement} This work has been partially supported by the network CNRS GDR 2932 "Th\'eorie des Jeux: Mod\'elisation math\'ematiques  et Applications" by the Commission of the
European Communities under the 7-th Framework Programme Marie
Curie Initial Training Networks   Project "Deterministic and
Stochastic Controlled Systems and Applications"
FP7-PEOPLE-2007-1-1-ITN, no. 213841-2 and   project SADCO ,
FP7-PEOPLE-2010-ITN, No 264735. This was also supported partially
by the French National Research Agency
 ANR-10-BLAN 0112.

\end{document}